\documentclass[12pt]{iopart}
\usepackage{braket} 
\usepackage{xcolor}
\usepackage{graphicx}
\usepackage{textcomp}
\usepackage{floatrow}
\usepackage{subfigure}
\usepackage{sidecap}
\usepackage{bm}

\newcommand{\red}[1]{\textcolor{red}{#1}}

\newcommand{\citespectro}{\cite{ZambriniCruzeiro2018a,Longdell2006,MA2018}}

\begin{document}

\title{Optical storage for 0.53 seconds in a solid-state atomic frequency comb memory using dynamical decoupling}

\author{Adrian Holz\"{a}pfel, Jean Etesse, Krzysztof T. Kaczmarek, Alexey Tiranov, Nicolas Gisin, Mikael Afzelius}

\address{Department of Applied Physics, University of Geneva, CH-1211 Geneva 4, Switzerland}
\ead{mikael.afzelius@unige.ch}
\vspace{10pt}
\begin{indented}
\item[] \today
\end{indented}

\begin{abstract}
Quantum memories with long storage times are key elements in long-distance quantum networks. The atomic frequency comb (AFC) memory in particular has shown great promise to fulfill this role, having demonstrated multimode capacity and spin-photon quantum correlations. However, the memory storage times have so-far been limited to about one millisecond, realized in a Eu${}^{3+}$ doped Y${}_2$SiO${}_5$ crystal at zero applied magnetic field. Motivated by studies showing increased spin coherence times under applied magnetic field, we developed a AFC spin-wave memory utilizing a weak 15 mT magnetic field in a specific direction that allows efficient optical and spin manipulation for AFC memory operations. With this field configuration the AFC spin-wave storage time increased to 40~ms using a simple spin-echo sequence. Furthermore, by applying dynamical decoupling techniques the spin-wave coherence time reaches 530~ms, a 300-fold increase with respect to previous AFC spin-wave storage experiments. This result paves the way towards long duration storage of quantum information in solid-state ensemble memories.
\end{abstract}

%
%
%
%
%

\section{Introduction}

A future quantum internet relies on the capability of remotely sharing quantum information, and last years have seen rapid progresses in increasing the distances over which it can be distributed. Recent experiments have demonstrated fiber-based quantum communication of over 400 km \cite{Boaron18,Lucamarini2018}, but reaching continental distances using fiber networks will require quantum repeaters \cite{Sangouard11}. These will require multiplexed quantum memories, i.e. devices that allow storage of quantum states of light in different modes in time, space or frequency \cite{Bussieres2013}. Another key feature is the ability to store multiplexed quantum states on timescales of at least several hundred milliseconds without a significant deterioration of storage efficiency over time \cite{Collins2007}.\medskip

Optical quantum memories based on spin-states in atomic ensembles have shown the potential to fulfill these requirements, both in laser-cooled alkali vapors \cite{Radnaev2010,Nicolas2014,Yang2016,Pu2017,Tian2017} and rare-earth (RE) ion doped crystals \cite{Longdell2005,Usmani10,Heinze2013,Ferguson2016,Seri2017,Laplane2017}. The AFC memory \cite{Afzelius09} realized in RE materials is particularly interesting for quantum repeaters, due to its high mode capacity \cite{Usmani10,Sinclair2014,Jobez2016,Laplane2016a}, high potential efficiency \cite{Sabooni2010,Jobez2014}, and demonstrated capability of storing quantum states \cite{Ferguson2016,Seri2017,Laplane2016a,Saglamyurek2011}. The storage time of spin-wave AFC memories in RE ion doped crystals, however, has been limited to a few milliseconds, realized in  Eu${}^{3+}$ doped Y${}_2$SiO${}_5$ crystals \cite{Laplane2017,Laplane2016a,Jobez15}. In these experiments the spin storage is realized on a zero-field nuclear quadrupole resonance where, at zero applied magnetic field, each quadrupole state is composed of two degenerate nuclear Zeeman states.\\
 In this article we show that by lifting the degeneracy using a weak external field, the AFC spin-storage time can be extended to 40~ms using a simple spin-echo sequence composed of two $\pi$ pulses. The direction of the magnetic field is carefully chosen based on previous spectroscopy studies \citespectro, in order to achieve efficient coherent optical and spin manipulation without cross talk between transitions. Furthermore, the spin-wave storage time can be extended to up to 0.53 s using a dynamical decoupling (DD) sequence \cite{viola1999dynamical}, two orders of magnitude longer than previous AFC storage experiments.\\ The article is organized as follows: in section 2, we remind the principle of the AFC protocol and the phenomena that impact its performances in terms of storage time, and describe the technique that we use to push the limitations further in our particular system. In section 3, experimental results are presented, and a spectral diffusion model is developed to explain the observed behaviors. Finally, discussions on discrepancies between the model and the experimental data are made in section 4.

\section{Spin-wave AFC and radio frequency manipulations}
\subsection{The protocol}
\label{subprotoclimit}
\begin{figure*}[ht!]
\includegraphics[width=0.4\textwidth]{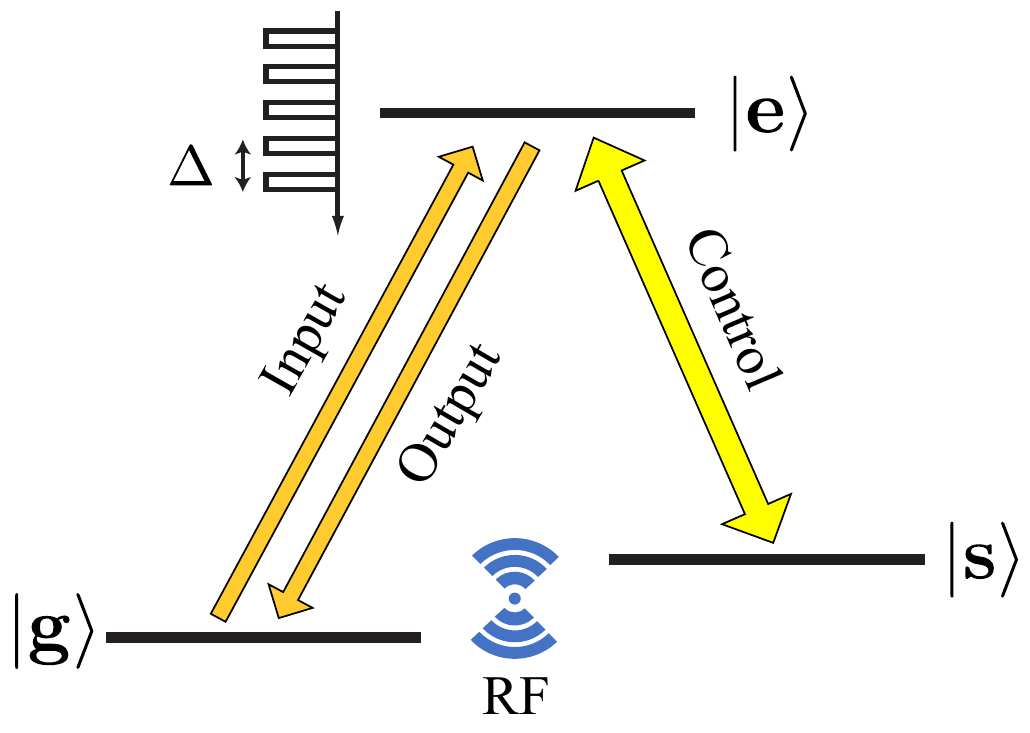}
\caption{\label{FIG:AFC} The spin-wave AFC protocol, in a lambda scheme.}
\end{figure*}

First, let us introduce the general principle of the AFC protocol, sketched in figure~\ref{FIG:AFC}. It is based on the creation of a frequency grating (the comb) on an optical transition $\ket{g}\leftrightarrow\ket{e}$ in an inhomogeneously broadened ensemble. The method we use for creating an AFC is detail in Ref. \cite{Jobez2016}. An input photon that is absorbed on this transition will create a single delocalized excitation, bringing the ensemble into a Dicke state:
\begin{equation}
\ket{\psi}\propto\sum_{j= 1}^Ne^{-i2\pi n_j\Delta t}\ket{g_1,...,e_j,...g_N},
\label{Dicke}
\end{equation}
where $n_j$ is the number of the comb tooth to which the atom $j$ belongs, and $\Delta$ is the comb's periodicity. The ensemble will then naturally undergo dephasing. However, due to the periodic comb structure, after a time $t=1/\Delta$, the atoms will all rephase and subsequently re-emit the photon as an echo (denoted output in figure~\ref{FIG:AFC}).
In order to allow for an increased storage time and on demand read out, the coherence is mapped onto the long-lived spin transition $\ket{g}\leftrightarrow\ket{s}$. This is done by applying a $\pi$ pulse (denoted control in figure 1) on the $\ket{e}\leftrightarrow\ket{s}$ transition before the re-emission has occurred. To retrieve the excitation, a second identical $\pi$ pulse is applied to map the coherence back onto the $\ket{g}\leftrightarrow\ket{e}$ transition where it rephases due to the AFC structure. In this article, we will note the duration between the two control pulses $T_{\rm spin}$, such that the total storage time is $T_{\rm spin}+1/\Delta$. The efficiency of the whole process can be expressed as:
\begin{equation}
\eta_{\rm tot}=\eta_{\rm AFC}(\eta_{\rm ctrl})^2\eta_{\rm spin},
\label{efftotsw}
\end{equation}
where $\eta_{\rm AFC}$ is the efficiency of the AFC protocol without spin storage, $\eta_{\rm ctrl}$ is the efficiency of a single control pulse and $\eta_{\rm spin}$ quantifies the degree to which coherence can be preserved while storing in the spin transition. There are two main mechanisms that contribute to loss of coherence during the spin storage, and consequently to a decrease of $\eta_{\rm spin}$.

The first mechanism stems from the inhomogeneous broadening of the spin transition $\Gamma_{\rm inh}$: as the different ions have slightly different resonance frequencies, they will dephase with respect to each other in a characteristic time given by $\sim1/\Gamma_{\rm inh}$. This effect can be undone by applying well known spin-echo techniques: two radio-frequency (RF) pulses are applied for performing $\pi$ rotations on the $\ket{g}\leftrightarrow\ket{s}$ transition (see figure~\ref{FIG:AFC}). Thanks to this technique, in \cite{Jobez15} we were able to push the storage time from tens of microseconds to milliseconds, while preserving a good signal-to-noise ratio at the single photon level.

The second mechanism comes from spectral diffusion, which is the variations of the spin transition frequencies over time, due to fluctuations in the ion's environment. This inevitably leads to a dynamical dephasing of the collective Dicke state (\ref{Dicke}), and thus to a decrease of the memory efficiency as a function of storage time.

The detrimental effect of spectral diffusion can be reduced by performing DD of the spin transition: a large number of $\pi$ pulses is applied on the $\ket{g}\leftrightarrow\ket{s}$ transition at a rate that is fast enough to decouple the spins from the fluctuations of the environment. In this case, the ions will spend as much time in the ground state $\ket{g}$ as in the excited state $\ket{s}$ of the spin transition, leading to a compensation of the dephasing and a longer effective coherence time. This spin echo technique has already been successfully applied to push the spin coherence times in RE ion doped crystals under a particular magnetic field condition, the ZEFOZ (ZEro First Order Zeeman) point, bringing the effective coherence time to up to six hours \cite{Zhong15}. Unfortunately, in Eu${}^{3+}$ doped Y${}_2$SiO${}_5$, reaching ZEFOZ points require magnetic field intensities of the order of 1~T \cite{Zhong15,Longdell2006}, making this configuration challenging to implement. We have chosen a different approach, for which DD is performed under a weak magnetic field. Previous studies have shown that the application of a weak field can also enhance the observed coherence time \cite{Equall1994}. In the following we will detail the different considerations that were made in choosing the magnetic field direction.\\

\subsection{Dynamical decoupling under magnetic field}
\label{subseq:DDtheo}
\begin{figure*}[ht!]
\includegraphics[width=\textwidth]{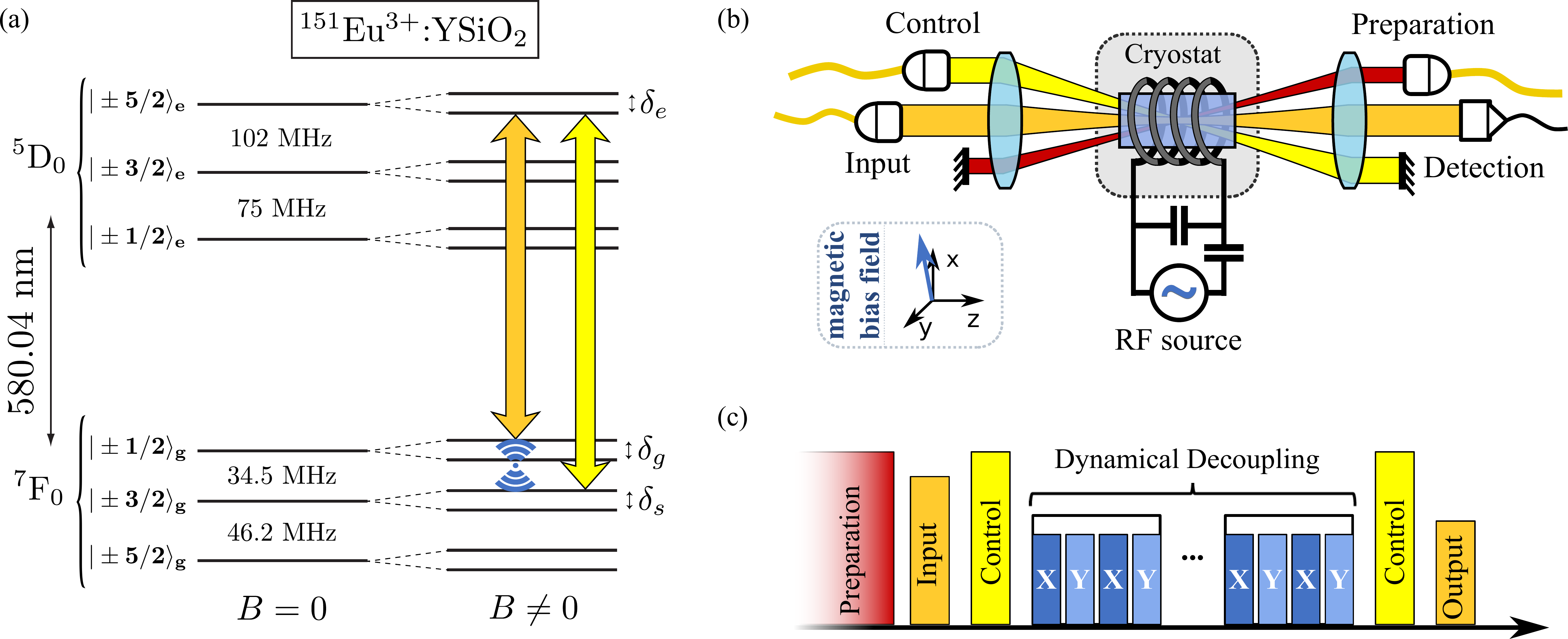}
\caption{\label{FIG:exp_scheme} The AFC protocol, and spin manipulation. (a) Lambda schemes for spin-wave AFC. The splits are all of the order of $\delta_i\simeq 10$~MHz/T, leadings to splits of $\delta_i\simeq 100$~kHz at 15 mT magnetic field. (b) Experimental setup used. (c) Time sequence for spin-wave AFC with DD.}
\end{figure*}

Our memory is based on a europium doped yttrium orthosilicate crystal ($^{151}$Eu$^{3+}$:Y$_2$SiO$_5$) in a non-zero magnetic field configuration. The level structure of this material is shown in figure~\ref{FIG:exp_scheme}(a), and consists of an optical transition between the electronic states ${}^7$F${}_0$ and ${}^5$D${}_0$ connecting two nuclear spin manifolds $I=5/2$ each of which can be described by the simplified Hamiltonian \cite{Teplov1968,Longdell2002}:
\begin{equation}
H_{\rm spin}=\mathbf{\hat{I}}\cdot \mathbf{Q}\cdot \mathbf{\hat{I}}+\mathbf{B}\cdot \mathbf{M}\cdot \mathbf{\hat{I}}.
\label{SpinHamilt}
\end{equation}
The effective-quadrupolar term $\mathbf{\hat{I}}\cdot \mathbf{Q}\cdot \mathbf{\hat{I}}$ is responsible for the zero-field energy level splittings of the order of $\sim$10~MHz that are shown in the right part of figure~\ref{FIG:exp_scheme}(a). Three levels are then at our disposal both in the optical ground and excited states, and already allowed us to implement the previously mentioned spin-wave AFC protocol by using states $\ket{g}=\ket{1/2}_g$, $\ket{s}=\ket{3/2}_g$ and $\ket{e}=\ket{5/2}_e$ \cite{Laplane2017,Laplane2016a,Jobez15}. As the splitting between $\ket{g}$ and $\ket{s}$ is 34.54 MHz, the spin manipulation could be performed by using a RF field, but unfortunately decays of the echos proved to be too short to apply dynamical decoupling techniques. Previous studies have found that the observed decay of the coherence can be enhanced by applying a weak magnetic field \cite{Equall1994}. In addition, such a field will obviously lift the degeneracy of each Zeeman doublet, due to the $\mathbf{B}\cdot \mathbf{M}\cdot \mathbf{\hat{I}}$ term in Eq.(\ref{SpinHamilt}).
In our case, the strength of the split is of the order of $\sim$10~MHz/T (see right part of figure \ref{FIG:exp_scheme}(a)).

We have decided to apply the magnetic field at $65^{\circ}$ relative to the D$_1$ axis in the plane spanned by the D$_1$ and D$_2$ polarization axis of the crystal \cite{Li1992}. Three major points motivated this choice of orientation: the reduced number of levels regarding site degeneracy, the branching ratio state-selection, and the equal ground states splittings $\delta_g=\delta_s:=\delta$ \citespectro\ (see notations of figure~\ref{FIG:exp_scheme}(a)). The precise considerations and implications for the choice of this field direction are found in the appendix; in short, operating with this bias field essentially allows us to profit from the increased spin coherence time while retaining the same optical depth as well as efficiency of spin control that we would have if no external field was applied. 

The strength of the magnetic field was set to the maximum possible value of the magnetic coils used for this experiments, which was 15 mT. At this field strength we note that:

\numparts
\begin{eqnarray}
\quad\delta=210~{\rm kHz}&>&\Gamma_{\rm inh}\label{condinhspin}\sim 30~{\rm kHz}\\
\quad\delta=210~{\rm kHz}&>&{\Omega_{\rm RF}}/{(2\pi)}\label{condrabi}=23~{\rm kHz}\\
\quad\delta_e=300~{\rm kHz}&>&\Gamma_{\rm AFC}\label{condBWAFC}=160~{\rm kHz}.
\end{eqnarray}
\endnumparts
On one hand, inequalities (\ref{condinhspin}) and (\ref{condrabi}) simply ensure that the RF field only excites the desired transition $\ket{+1/2}_g\leftrightarrow\ket{+3/2}_g$ or $\ket{-1/2}_g\leftrightarrow\ket{-3/2}_g$ without exciting the crossed transitions $\ket{+1/2}_g\leftrightarrow\ket{-3/2}_g$ or $\ket{+1/2}_g\leftrightarrow\ket{-3/2}_g$. Inequality (\ref{condBWAFC}), on the other hand, puts a limit onto the AFC bandwidth $\Gamma_{\rm AFC}$  which stems from the AFC preparation and ensures that no additional optical depth is lost in the comb preparation process.\\

The AFC bandwidth is limited here by the strength of the applied magnetic field, which in turn limits the shortest possible input pulse duration and hence the temporal multimode capacity \cite{Afzelius09}. However, by using a stronger field of about 100 mT one can recover the AFC bandwidth that can be achieved at zero applied field. It should be noted that such fields are more readily produced in the lab with respect to the $>$1~T fields required to work at a ZEFOZ point in Eu${}^{3+}$ doped Y${}_2$SiO${}_5$.

\subsection{Experimental implementation}

The experimental setup is shown in figure~\ref{FIG:exp_scheme}(b). The present AFC spin-wave memory is implemented in a 1000 ppm isotopically pure $^{151}$Eu$^{3+}$ doped Y$_2$SiO$_5$ crystal (peak optical absorption $\alpha=2.6$~cm${}^{-1}$, optical inhomogeneous broadening $\Gamma_{\rm inh}^{\rm opt}\sim1.5$~GHz), which is cooled down to 4K in a closed-cycle helium cryostat. The crystal is attached to a custom-made vibration damping mount, in order to allow for the preparation of spectrally narrow structures with the cryostat running \cite{ZambriniCruzeiro2018a,Louchet_Chauvet_2019}. The vacuum chamber that contains the crystal is surrounded by three pairs of coils in Helmholtz configuration, allowing us to apply magnetic bias fields in an adjustable direction. To fine tune this direction to the desired $65^{\circ}$ relative to D$_1$, we use spectral hole burning techniques as described in \cite{ZambriniCruzeiro2018a}. At this angle and with a field intensity of 15~mT, the ground state splittings are both equal to $\delta = 210 $~kHz and the excited state splitting is $\delta_e=300$~kHz. Given that the spin inhomogeneous broadening of our crystal is of the order of $\Gamma_{\rm inh}\sim30$~kHz, this field intensity allows us to fulfill condition (\ref{condinhspin}). To optically address the ions we use a laser at 580.04 nm that is reference locked to another laser stabilized on a high finesse cavity. From spectral hole burning experiments we estimate its linewidth to less than 1~kHz.  The laser is split into three optical spatial modes, as shown in figure~\ref{FIG:exp_scheme}(b): the input (orange in the figure, with a waist $w_0^{\rm in}=33$~\textmu m in the crystal), the control (yellow in the figure, $w_0^{\rm ctr}=350$~\textmu m) and the preparation mode (red in the figure, $w_0^{\rm prep}=480$~\textmu m). The different modes are used because they allow us to reduce detection noise through leakage and because we observe a different optimum for AFC preparation and control mode regarding the trade-off between intensity and homogeneity over the storage volume. All optical modes propagate close to the b axis of the crystal, i.e. orthogonal to the D$_1$ and D$_2$ axis, and are polarized along D$_1$ to maximize absorption \cite{Koenz2003,Ferrier2016}.\\

The experimental sequence is illustrated in figure~\ref{FIG:exp_scheme}(c). We prepare an AFC structure of inverse periodicity $1/\Delta=17$~\textmu s over a bandwidth of $\Gamma_{\rm AFC}=160$~kHz (which satisfies condition (\ref{condBWAFC})), using the preparation mode and a preparation scheme as described in \cite{Jobez_2014}. The input field is a classical coherent state (Gaussian, FWHM = 7~\textmu s) and the control fields are intense Fourier-limited $\pi$-pulses (square, FWHM = 4~\textmu s). For the RF control we use a simple resonant LC circuit (Q-factor Q = 25) consisting of a coil wrapped around the crystal (inductance), a parallel and a serial capacitor, as sketched in figure~\ref{FIG:exp_scheme}(b). The resonance can easily be tuned to 34.54~MHz by adjusting the two capacitances placed outside of the cryostat. With an input RF power of several tens of watts we are able to drive the spin transition with a Rabi frequency $\mathrm{\Omega_{RF}}$~=~$2\pi\times$23~kHz. As we want to address ions efficiently over the whole inhomogeneous broadening $\Gamma_{\rm inh}$, quasi-monochromatic pulses would be insufficient given our limited Rabi frequency. Fortunately, efficient $\pi$ rotations over the entire spin linewidth can be achieved by using adiabatic pulses \cite{Rippe2005}. Here we used hyperbolic secant profiles with FWHM of 80~\textmu s and a chirp in frequency of 60~kHz.

\section{Experimental results}

\begin{figure*}[ht!]
\includegraphics[width=1\textwidth]{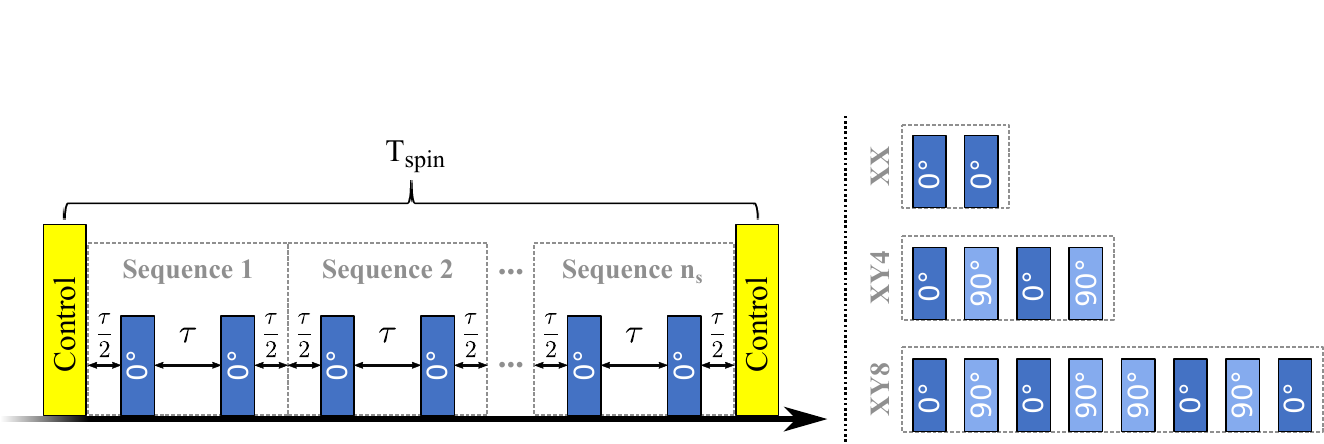}
\caption{\label{FIG:DDdetail} \textbf{left:} Illustration of a DD sequence for the case of decoupling with an XX-sequence. \textbf{right:} Phase-relations between the decoupling pulses for the three different sequences that we implemented in this experiment.}
\end{figure*}

In order to characterize DD in our system, we performed the storage experiment as described previously while varying several different parameters of our decoupling scheme. The different parameters are illustrated in figure \ref{FIG:DDdetail}. In general, we perform the decoupling by applying $n_s$ sequences, each consisting of an even number of $\pi$-pulses $n_p$. Between adjacent $\pi$-pulses there is a delay $\tau$, such that we apply a total  of $n=n_sn_p$ pulses over an overall spin storage time of $T_{\rm spin}=n\tau$.

\subsection{Storage experiments with fixed number of pulses}
\label{subseq:OU}

\begin{figure*}[ht!]
	\includegraphics[width=0.6\textwidth]{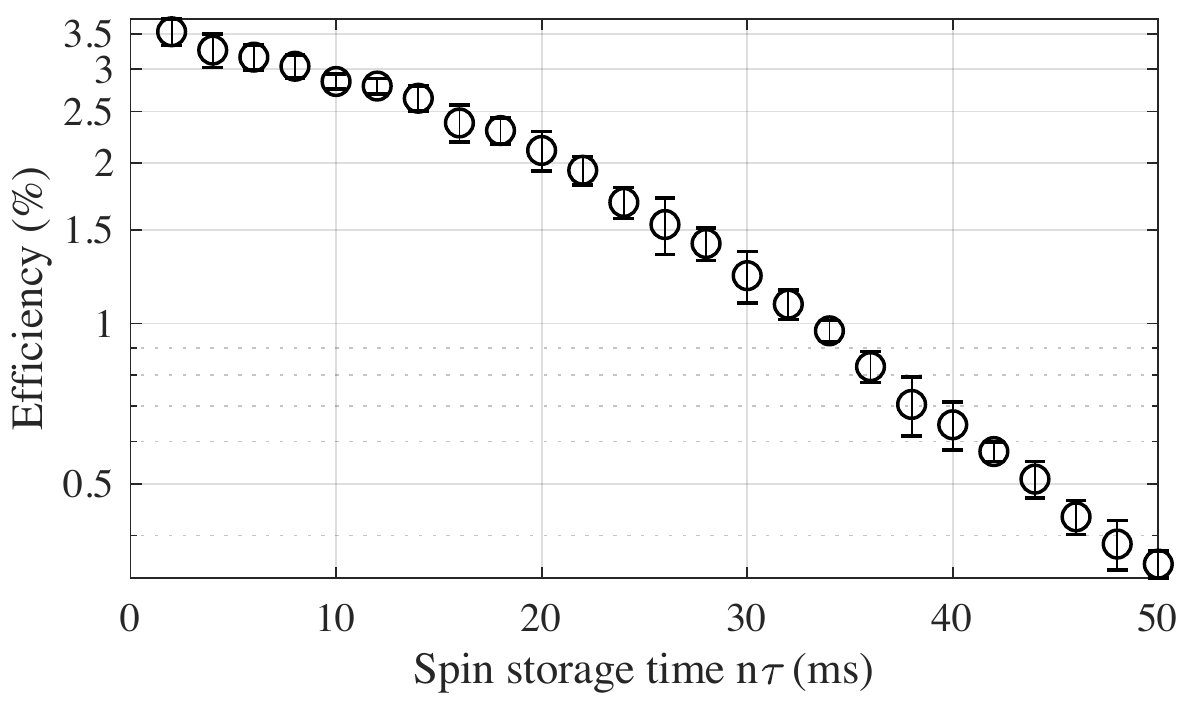}
	\caption{ Efficiency as a function of storage time with two identical rephasing $\pi$-pulses ($n_s=1$, $n_p$ = 2, such that $n=2$). For the shortest spin storage time of $T_{\rm spin}=n\tau=$~=~2~ms the memory efficiency was $\eta_{\rm tot}~=~3.7\pm0.2\%$.}
	\label{FIG:decay_curve_2pulses}
\end{figure*}

We start by characterizing the AFC spin-wave memory performance using a single sequence consisting of two identical $\pi$-pulses ($n_s=1$, $n_p=2$), which is the minimum number of required pulses in an optical storage experiment in order to compensate the inhomogeneous broadening of the spin transition (see e.g. \cite{Longdell2005,Jobez15}). The memory efficiency was measured while changing $\tau$. In figure~\ref{FIG:decay_curve_2pulses} the resulting memory efficiency is plotted as a function of $T_{\rm spin}=n\tau$. For the shortest storage time ($T_{\rm spin}$~=~2~ms), the efficiency reaches $\eta_{\rm tot}~=~3.7\pm0.2\%$. Taking into account the AFC and the control efficiencies $\eta_{\rm AFC}~=~10.2\pm 0.7\%$ and $\eta_{\rm ctrl} = 61\pm 2\%$ (both measured with independent methods), Eq.(\ref{efftotsw}) suggests that the spin wave efficiency $\eta_{\rm spin}$ is close to unity, within the error. Note that $\eta_{\rm spin}$ takes into account both population transfer errors and phase coherence errors due to the spin manipulation. These results indicate that the ions are efficiently manipulated over the whole inhomogeneous spin linewidth, and that no unaccounted experimental inefficiency affects the memory performance. We estimate the population transfer error of individual $\pi$-pulses to be about 2\%, see the discussion in section \ref{subseq:tau_dep}.

\begin{figure*}[ht!]
	\subfigure[]{
		\includegraphics[width=0.46\textwidth]{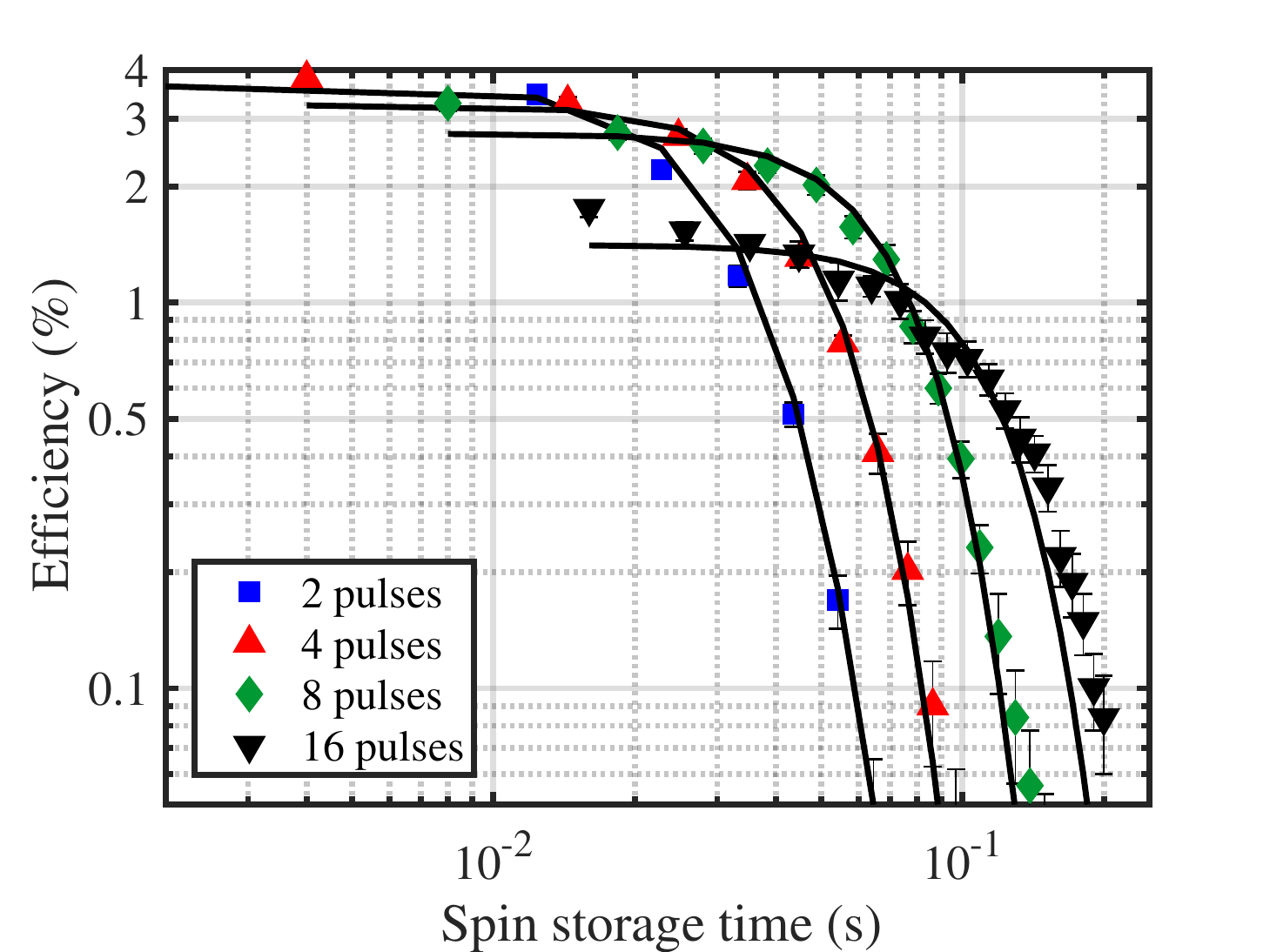}}
	\subfigure[]{
		\includegraphics[width=0.46\textwidth]{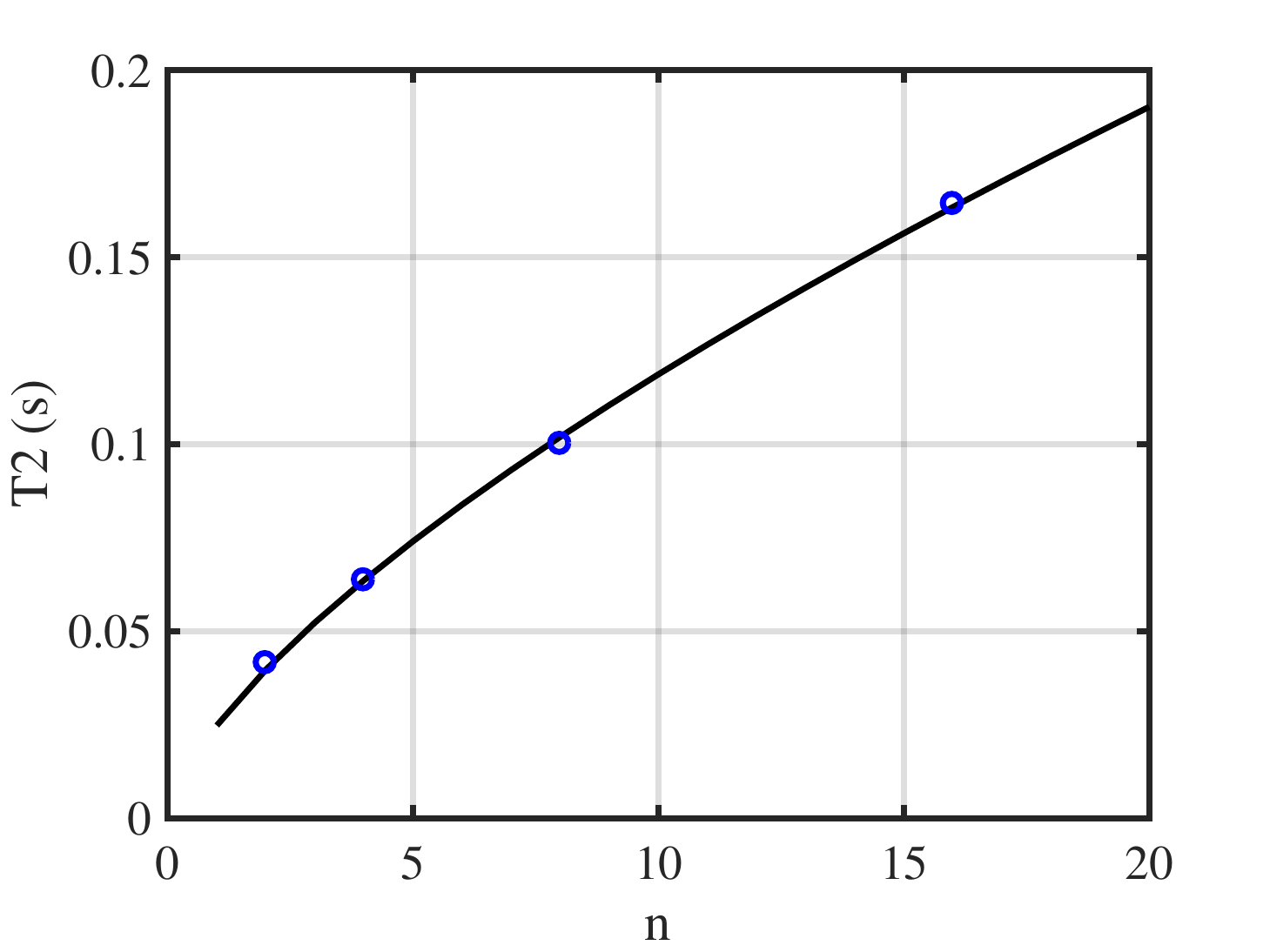}}
	\caption{\label{FIG:fixedpulsenumber}AFC spin wave storage with a fixed number of rephasing pulses. (a) Efficiency as a function of storage time for different numbers $n$ of identical $\pi$-pulses acting on the spins, for $n = 2, 4, 8, 16$. The solid lines represent the fitted OU spectral diffusion model of eq. (\ref{OUeq_general}) as discussed in the text. (b) Characteristic time scales of the decay curves, as a function of number of pulses $n$. The solid-line shows the fit to a theoretical power law, see text for details.}
\end{figure*}

The non-exponential shape of the efficiency curve shown in figure~\ref{FIG:decay_curve_2pulses} indicates decoherence due to spectral diffusion \cite{Klauder1962, Mims1968} that might be mitigated by DD techniques \cite{viola1999dynamical,Zhong15,Pascual-Winter2012, Arcangeli2014, Heinze2014}. To demonstrate that DD can indeed extend the coherence time we performed storage experiments with increasing number of sequences $n_s = 1, 2, 4, 8$ while maintaining $n_p=2$ (the global number of pulses is then $n=2,4,8,16$). As in figure~\ref{FIG:decay_curve_2pulses} the number of pulses were kept constant for each experiment and the pulse separation $\tau$ was varied. As seen in figure~\ref{FIG:fixedpulsenumber}(a), one clearly observes an increase of the storage time as the number of pulses is increased, which indicates that DD is effective in reducing the rate of dephasing due to spectral diffusion. Following Ref. \cite{Medford2012}, we fit all curves to a stretched exponential (SE):
\begin{equation}
\eta^{\rm SE}_{\rm spin}(n,\tau)=\exp\left[-2\left(\frac{n\tau}{T_2(n)}\right)^\alpha\right].
\label{EQ:Mims}
\end{equation}
We underline here that this particular form only aims at extracting a characteristic time $T_2(n)$ for the decay, which only loosely depends on $\alpha$ \cite{Medford2012}. The resulting $T_2(n)$ as a function of the number of pulses $n$ is shown in figure~\ref{FIG:fixedpulsenumber}(b), and it follows a power-law scaling $T_2(n) =T_2(1) n^{\gamma}$ where $T_2(1)=25\pm1$~ms and $\gamma = 0.68 \pm 0.02$. This value of $\gamma$ is actually close to the one that we expect in the case of spectral diffusion governed by an Ornstein-Uhlenbeck (OU) process \cite{Klauder1962}, for which $\gamma^{\rm OU} = 2/3$ \cite{Medford2012,Sousa2009, Lange2010}. In this model, the detuning of each spin diffuses in a Markovian fashion within a characteristic time $\tau_c$ into a Gaussian steady state distribution of spectral width $\sigma$, leading to a dephasing rate of\footnote{Note that this expression has been slightly approximated with respect to Eq. (A15) in \cite{Pascual-Winter2012}} \cite{Pascual-Winter2012}:

\numparts
\begin{eqnarray}
\eta^{\rm OU}_{\rm spin}(n,\tau)=&\exp\left[{-2\Gamma(n,\tau)}\right],\quad \rm{with}\label{OUeq_general1}\\
\Gamma(n,\tau) =& (\sigma \tau_c)^2 \left(\left[ \frac{1}{\tau_c} - \frac{2}{\tau} \tanh\left(\frac{\tau}{2\tau_c}\right)\right] n \tau - \left[ 1-\rm sech \left( \frac{\tau}{2\tau_c} \right)  \right]^2 \right).
\label{OUeq_general}
\end{eqnarray}
\endnumparts
\noindent To gain a more quantitative understanding of the dephasing process, the decay curves were then fitted to this model with $\sigma$ and $\tau_c$ as global, free parameters, and the results are shown as solid lines in figure~\ref{FIG:fixedpulsenumber}(a). The model fit is good for low $n$, but does not fit as well the data for high $n$, particularly for $n = 16$. As will be discussed later we suspect this to be caused by some technical error appearing for short pulse separations. However, overall the model fit is rather satisfactory and yields the spectral diffusion OU parameters $\sigma/(2\pi)=15.1~\pm3.5~\textrm{Hz}$ and $\tau_c=9.5~\pm1.2$~ms. Interestingly, given the gradient of the spin transition frequency $S_1\simeq17$~MHz/T \citespectro\ the equivalent fluctuation of the magnetic field at the position of the europium ion is of the order of $\Delta B=\sigma/S_1\sim$ 1 \textmu T, which is of the same order of magnitude as what has been found in other studies \cite{Zhong15,Fraval2004}.
We finally note that, knowing $\sigma$ and $\tau_c$, one can theoretically calculate the $T_2(1)$ value appearing in the power-law discussed in the previous section. Indeed, in the limit $\tau<<\tau_c$, (\ref{OUeq_general1}) simplifies to:
\begin{equation}
\eta^{\rm OU}_{\rm spin}(n,\tau)\simeq\exp\left[-\frac{\sigma^2\tau^3n}{6\tau_c}\right],
\label{OUsimple}
\end{equation}
leading to a theoretical dependance of $T_2(n) =T_2(1)n^{\gamma}$ with $\gamma=2/3$ and $T_2(1) =  \sqrt[3]{12\tau_c/\sigma^2} \approx 23$~ms \cite{Lange2010}, in good agreement with the value obtained using the power-law fit above.

\subsection{Storage experiments with fixed pulse separation}
\label{subseq:tau_dep}

In an optical storage application experiment where on-demand readout of the memory is to be performed, for instance conditioned on some external signal, then it is preferable to keep the pulse separation $\tau$ constant. In the case of ideal $\pi$-pulses, working with the smallest possible $\tau$ would yield the best decoupling effects (see (\ref{OUsimple})). In practice, however, the application of many inversion pulses introduces errors as the RF pulses do not perform perfect inversions \cite{Souza2011,Wang2012b,ZambriniCruzeiro2016}. This is already seen in figure~\ref{FIG:fixedpulsenumber}(\red{a}), where the efficiency when applying $n =16$ is clearly lower for short storage times, although a longer overall memory time is reached. Therefore, we expect a maximum effective coherence time of the memory device for some finite optimal value of the pulse separation. Furthermore, in addition to using sequences of identical RF pulses, known as a Carr-Purcell or XX sequence \cite{CarrPurcell1954}, we also studied more complex sequences that are more resilient against imperfect inversion pulses. In particular we investigated the sequences XY4 and  XY8, as described in figure \ref{FIG:DDdetail},  and determined the optimal value of $\tau$ for each of them.

For each storage experiment with fixed $\tau$ the memory efficiency was measured while increasing the number of pulses $n$. As all recorded decay curves were exponential within experimental errors, each of them was fitted to an exponential function $\eta^{DD}_{\rm spin}=\exp(-2 n \tau/T_2^{DD})$, allowing us to extract the effective coherence time $T_2^{DD}$ of the memory. Note that an exponential decay is expected from the OU spectral diffusion model, see (\ref{OUeq_general}), (\ref{OUsimple}) when $\tau$ is kept constant while $n$ is varied.

\begin{figure*}[ht!]
\subfigure[]{
\includegraphics[width=0.46\textwidth]{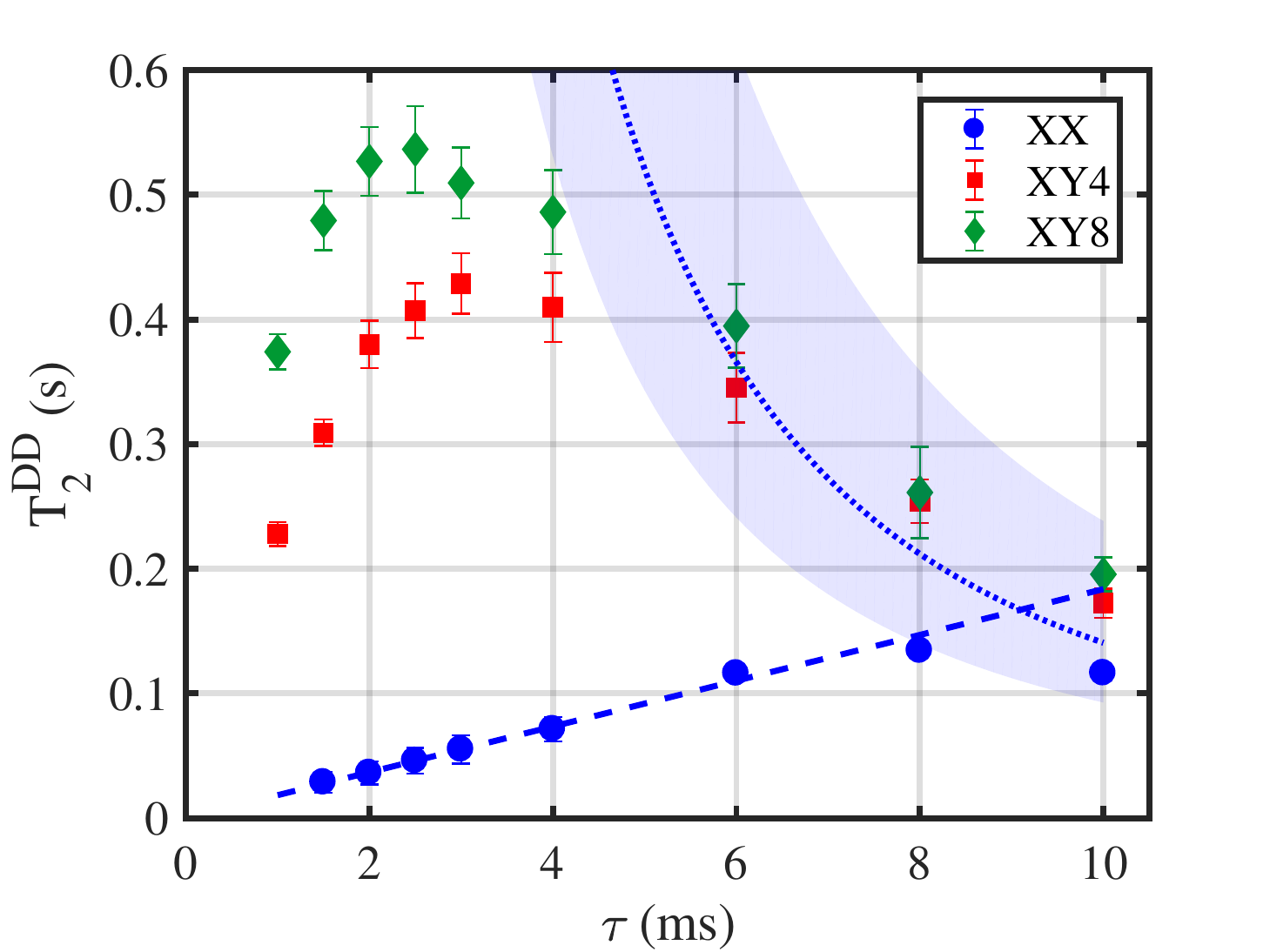}}
\subfigure[]{
\includegraphics[width=0.46\textwidth]{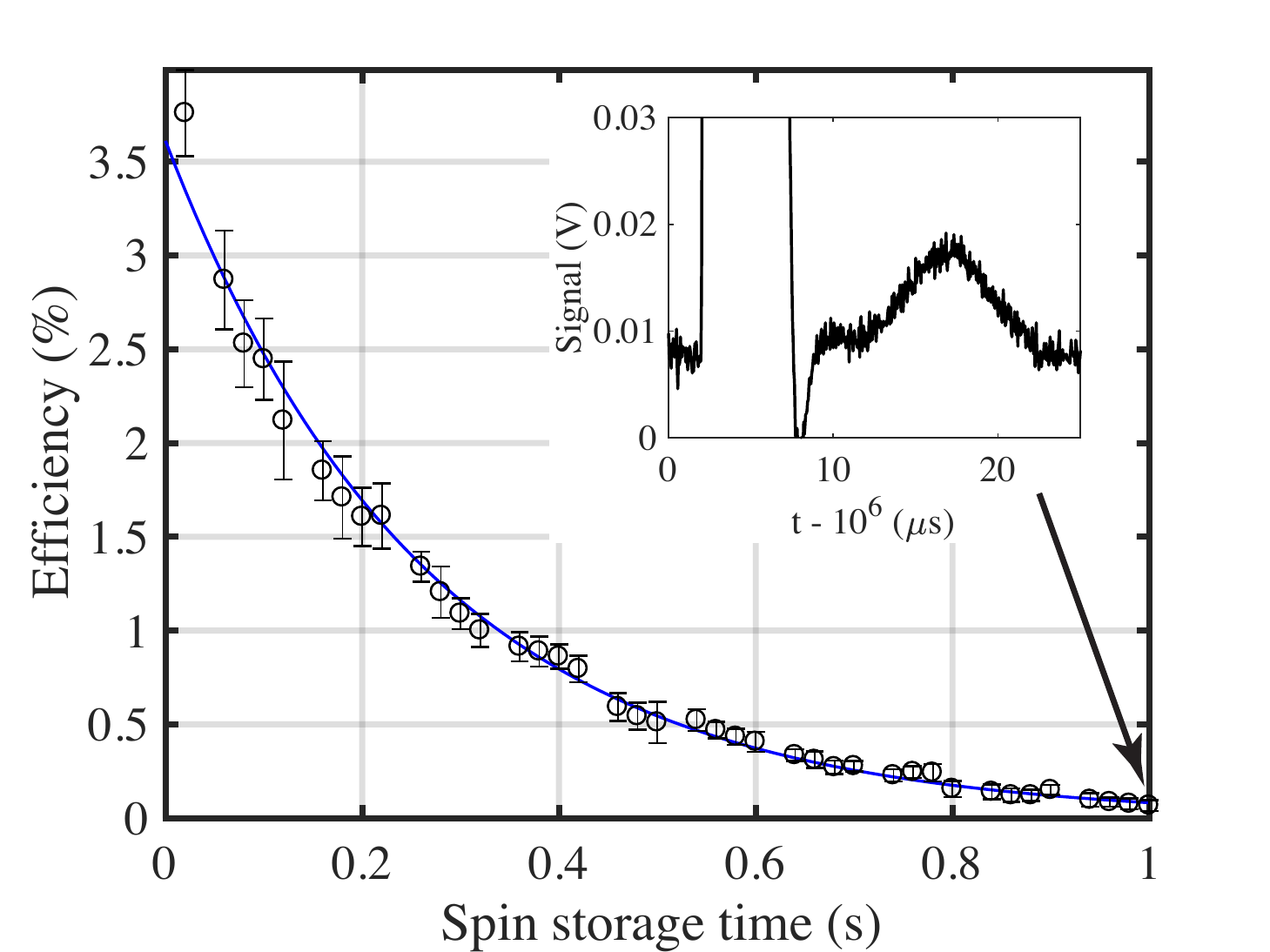}}
\caption{\label{FIG:DD}AFC spin-wave storage using different DD sequences with constant duration between RF inversion pulses. (a) Effective memory coherence time $T_2^{DD}$ as a function of decoupling pulse spacing $\mathrm{\tau}$, for DD pulse sequences XX, XY-4 and XY-8. Solid and dashed line come from modeling explained in the main text. (b) The memory efficiency decay curve for the longest achieved memory time of $T_2^{DD} = 0.53$~s using the XY8 sequence with $\mathrm{\tau} = 2.5$~ms.}
\end{figure*}

The experimental results of the effective memory coherence measurements are presented in figure~\ref{FIG:DD}(a). As expected the coherence time reaches a maximum value for an optimum pulse separation, and the use of more error-resilient pulse sequences allows one to use shorter pulse separation which results in longer coherence times. The longest coherence time was achieved using a XY8 sequence with $\tau$ = 2.5~ms, and the corresponding efficiency curve is shown in figure~\ref{FIG:DD}(b). The resulting coherence time of $0.53~\textrm{s}~\pm~0.03~\textrm{s}$ represents a more than 10-fold increase compared to storage without decoupling, see figure~\ref{FIG:fixedpulsenumber}(a), and a 300-fold increase compared to the previous state of the art in AFC spin-wave storage \cite{Laplane2017,Laplane2016a,Jobez15}. The echo at one second is also shown in inset of figure~\ref{FIG:DD}(b) and shows that even at this delay it can be well discriminated. 

In order to analyse the effective coherence time more quantitatively, we use two models that describe the observed dependence in two complementary, asymptotic regimes. For short pulse separation $\tau$ we expect coherence time to be predominantly limited by the inversion error that the pulses introduce into our system. For long pulse separations, i.e. when there are few inversion pulses applied, the error that is introduced by the pulses should be negligible compared to the effect of spectral diffusion, which we model using the OU diffusion model.

The pulse inversion error is modeled using the theory presented in Ref. \cite{ZambriniCruzeiro2016}, which considered an ideal $\pi$ inversion pulse with an area error $\epsilon$ (pulse area $\theta = \pi + \epsilon$). With this model, we expect an effective coherence time $T_{2,\epsilon}^{DD} = \sqrt{2/\alpha} n_p \tau$. The pulse error dependence enters into the $\alpha$ parameter, which is $\alpha = \epsilon^2$,$\epsilon^4/2$ and $\epsilon^6/4$ for the XX,XY-4 and XY-8 sequences \cite{ZambriniCruzeiro2016}. We note that this model predicts a Gaussian time dependence of the decay. The decays recorded using the XX sequence showed a tendency of Gaussian decay for short pulse separations, and exponential decays for longer pulse separations, see the appendix. However, the mean difference in the fitted coherence time using either a Gaussian or exponential decay was less than 6$\%$ (see appendix for a more detailed discussion). To simplify the discussion here all decay curves were fitted with exponential decays.

According to the pulse area error model the coherence time should increase linearly with the pulse separation, as already noted in Ref. \cite{ZambriniCruzeiro2016}. The experimental data also show approximately linear dependence at short pulse separations (see figure~\ref{FIG:DD}(a)), most clearly for the XX sequence. The first four points of the XX curve were fitted to the model curve $T_{2,\epsilon}^{DD} = 2\sqrt{2}\tau/\epsilon $, shown as a dashed line in figure~\ref{FIG:DD}(a), which resulted in $\epsilon = 0.154 \pm 0.004$~rad. This value is consistent with other measurements of the population error that we have performed, and close to the error previously measured in a similar RF set-up \cite{Jobez15}. 

The XY-4 and XY-8 coherence times also follow an approximately linear dependence for the shortest pulse separations, although the few number of points do not allow a quantitative comparison with the model. However, given the $1/\epsilon^2$ (XY-4) and $1/\epsilon^3$ (XY-8) scaling of the coherence time, significantly longer coherence times should be observed for short $\tau$. For instance, based on the $\epsilon= 0.154$ value fitted to the XX sequence, we would expect a XY-4 coherence time of about 340 ms for $\tau$~=~1~ms, while the experiment yielded about 230 ms. The discrepancy is even more significant for the XY-8 sequence. In summary, the coherence time at short delays can be enhanced by using error-compensating DD sequences, however, it appears that an additional factor limits the achievable coherence time. This could be due to errors not accounted for in the model, or due to some other dephasing process that does not follow a OU spectral diffusion model, see also the discussion in section \ref{sec_discussion}.

In the opposite regime of long pulse separations, where the number of $\pi$ pulses is smaller, we expect the spectral diffusion to limit the the achievable coherence time. The OU model predicts an exponential decay with an effective coherence time $T_{2,OU}^{DD} = 12\tau_c/(\sigma^2\tau^2)$, see eq. (\ref{OUsimple}). This model, with the OU parameters we extracted in section \ref{subseq:OU}, results in the dotted line in figure~\ref{FIG:DD}(a), which is in reasonably good agreement with the observed coherence times for the XY-4 and XY-8 sequences at large pulse separations where the effect of pulse errors is negligible. We note that it is strongly dependent on the $\sigma$ parameter, as shown by the blue shaded area in figure~\ref{FIG:DD}(a) which represents the prediction within the $\pm$~3.5~Hz one standard deviation error.

\section{Discussion}
\label{sec_discussion}
The presented results show the applicability and the effectiveness of the DD sequences to spin-wave AFC protocols in rare-earth ion doped crystals. In the course of its implementation we uncovered some open questions.\\
The first is related to the error-resistant decoupling sequences. As explained in the previous section, we find no angular error for the RF rephasing pulses that is both consistent with the observed improvement from XX to XY4 as well as the improvement from XY4 to XY8. This might indicate that our sequences suffer from more complex errors like phase errors or global errors due to heating. This could mean that cross-talk between the crossed RF transitions might still be present even if conditions (\ref{condinhspin}) and (\ref{condrabi}) are fulfilled. A way to minimize this contribution would be to work at even higher magnetic field amplitudes in order to push the crossed transitions further away in frequency. This problem should be solved first if we want to aim at more complex sequences like KDD \cite{Souza2011}.\\
The second open question is related to the model itself. Indeed, the OU process seems to describe the decoherence process in our system quite well, but there are a few discrepancies that suggest that not all our assumptions about the system are fulfilled. In particular, as we mentioned previously, the fit of the OU model for the four curves of figure~\ref{FIG:fixedpulsenumber}(a) cannot be adjusted such that low and high $n$ values can be well fitted. Further, even if we just consider the isolated 16 pulse curve, we do not find a set of parameters that describes both the behavior in the short $\tau$ as well as the long $\tau$ regime well. This might either signify that the pulse density affects their efficiency, as it would be true for heating effects of the RF circuit, or it might suggest that the OU process is not the only form of decoherence that our system experiences. Another process, for example, could be the coupling of the spins to some electromagnetic background AC field in the laboratory environment. 
Even in the presence of these discrepancies, we note that the OU model allows for a globally satisfying description of our system, and gives valuable information about the ion and his environment, that are in good agreement with previous studies.
\\

A future important step would be to characterize any additional noise that is introduced by the decoupling process. While we are confident that we can store and retrieve classical light pulses without any major background contribution, DD with imperfect pulses is expected to introduce noise that is not scaling with the amplitude of the input \cite{ZambriniCruzeiro2016}. Such noise might be a notable limitation in the storage at the single photon level, and would deserve further investigation. Another point that should also be addressed is the efficiency of the memory, currently limited to a few percent. Two main limitations play a role here. The first is the low optical depth of the memory. A common method to increase it is letting the input pass several times through crystal. Such a multi-pass configuration, however, requires interferometric stability over the whole duration of storage, which, in the presence of the vibrating cryostat, is technically difficult to achieve. An elegant way of bypassing this additional experimental complexity might be the use of a cavity within the crystal itself \cite{Sabooni2013}. The second limitation for the global efficiency is the control pulse transfer efficiency, that could be drastically increased by using waveguide designs for the memory \cite{seri2017quantum}.

\section{Conclusion and outlook}
We have presented and experimentally shown the application of a dynamic decoupling sequence under a weak magnetic field to the spin-wave atomic frequency comb protocol. A spectral diffusion model based on the Ornstein-Uhlenbeck process has proven to explain our experimental observations with good accuracy, and with parameters that are in good agreement with previous studies. Robust spin echo sequences have then been used to demonstrate storage over durations of the order of a second, with a characteristic decay time of more than 0.5 s. As next steps, thorough study of noise induced by the dynamical decoupling sequence \cite{Jobez15}, and efforts towards increasing the efficiency of these memories would allow significant advance in the development of efficient long lived quantum memories.

\section{Acknowledgments}
We would like to thank Claudio Barreiro for technical support. This work was financially supported by the European Union via the Quantum Flagship project QIA (GA No. 820445), and by the Swiss FNS Research Project No. 172590.
\appendix
\section{Considerations regarding the static magnetic bias field}
In the present appendix, we detail the different considerations that we have made to choose the direction and the magnitude of the magnetic field that we apply to our crystal, and discuss the resultant constraints on our experiment.
	
	\subsection{Choosing the direction of the field}
	There are two magnetically inequivalent sub-sites of Europium in Y$_2$SiO$_5$ that we are addressing in this experiment. These two subsites
behave the same under an external magnetic bias field, only when the latter is oriented along the $b$ symmetry axis of the crystal and in the plane orthogonal to this symmetry axis - the plane that is spanned by the D$_1$ and D$_2$ polarisation axis of the crystal \cite{Li1992}. We restrict ourselves to this particular plane to avoid further complications caused by the sub-sites.\\

	\begin{SCfigure}[50][ht!]
		\includegraphics[width=0.5\textwidth]{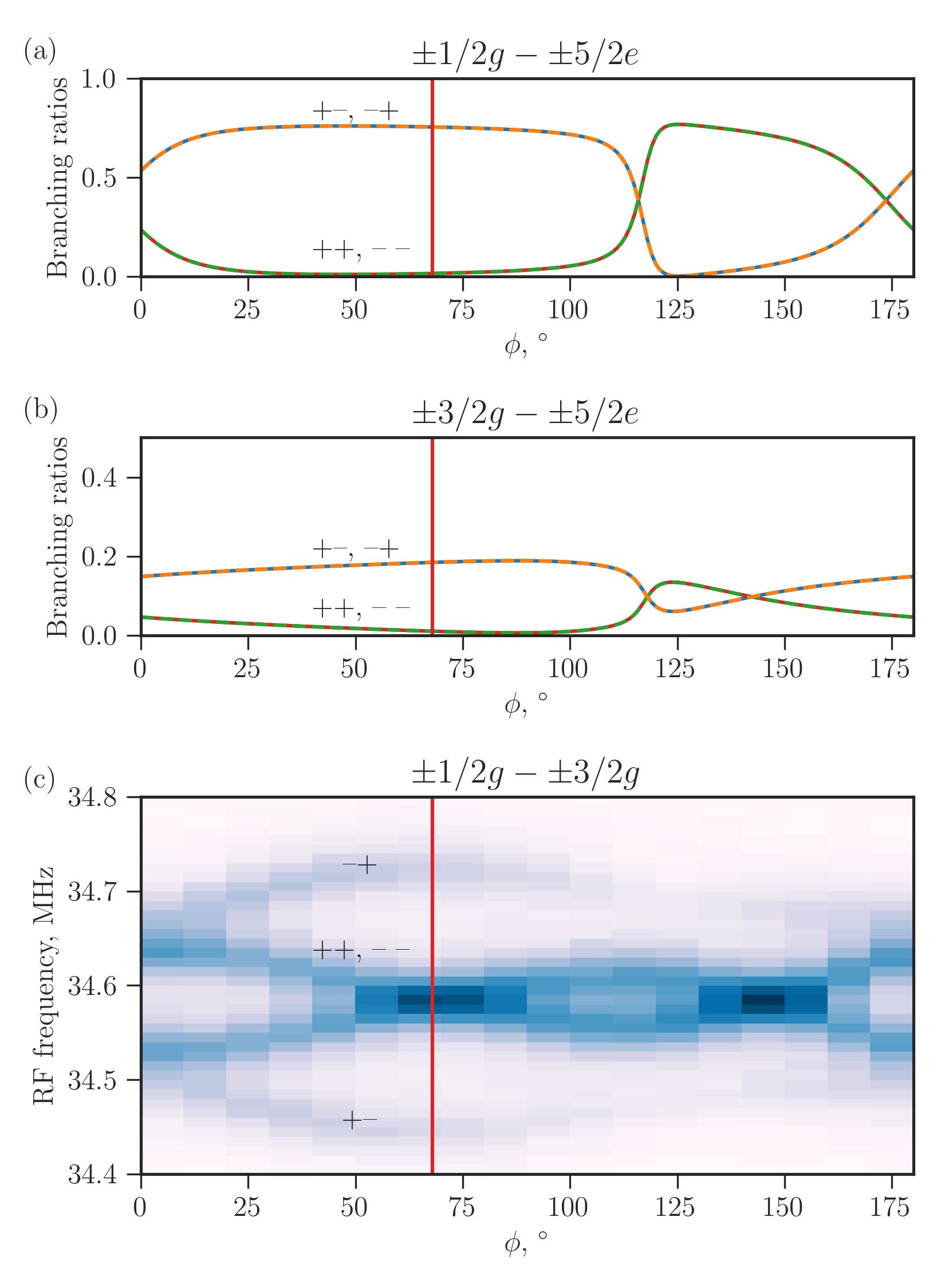}
		\caption{\label{FIG:branchingratios}\textbf{(a)},\textbf{(b)} Branching ratios for transitions between states $\ket{5/2}_e$ and, respectively, $\ket{1/2}_g$ and $\ket{3/2}_g$ as a function of the angle of the applied magnetic field relative to the D$_1$-axis in the D$_1$-D$_2$-plane. For both cases at an angle of 65$^\circ$ (red line) transitions, where the sign of the quantum number changes, are strongly preferred. \newline
			\textbf{(c)} Transition strength and frequency dependence of the transitions $\ket{1/2}_g\leftrightarrow\ket{3/2}_g$ as a function of the angle of the magnetic bias field. At 65$^\circ$ two of the transitions coincide in frequency and have a strong dipole moment, while the other two are spectrally well-separated and have a weaker dipole moment.
		}	
	\end{SCfigure}
	
\begin{SCfigure}[50][ht!]
		\includegraphics[width=0.5\textwidth]{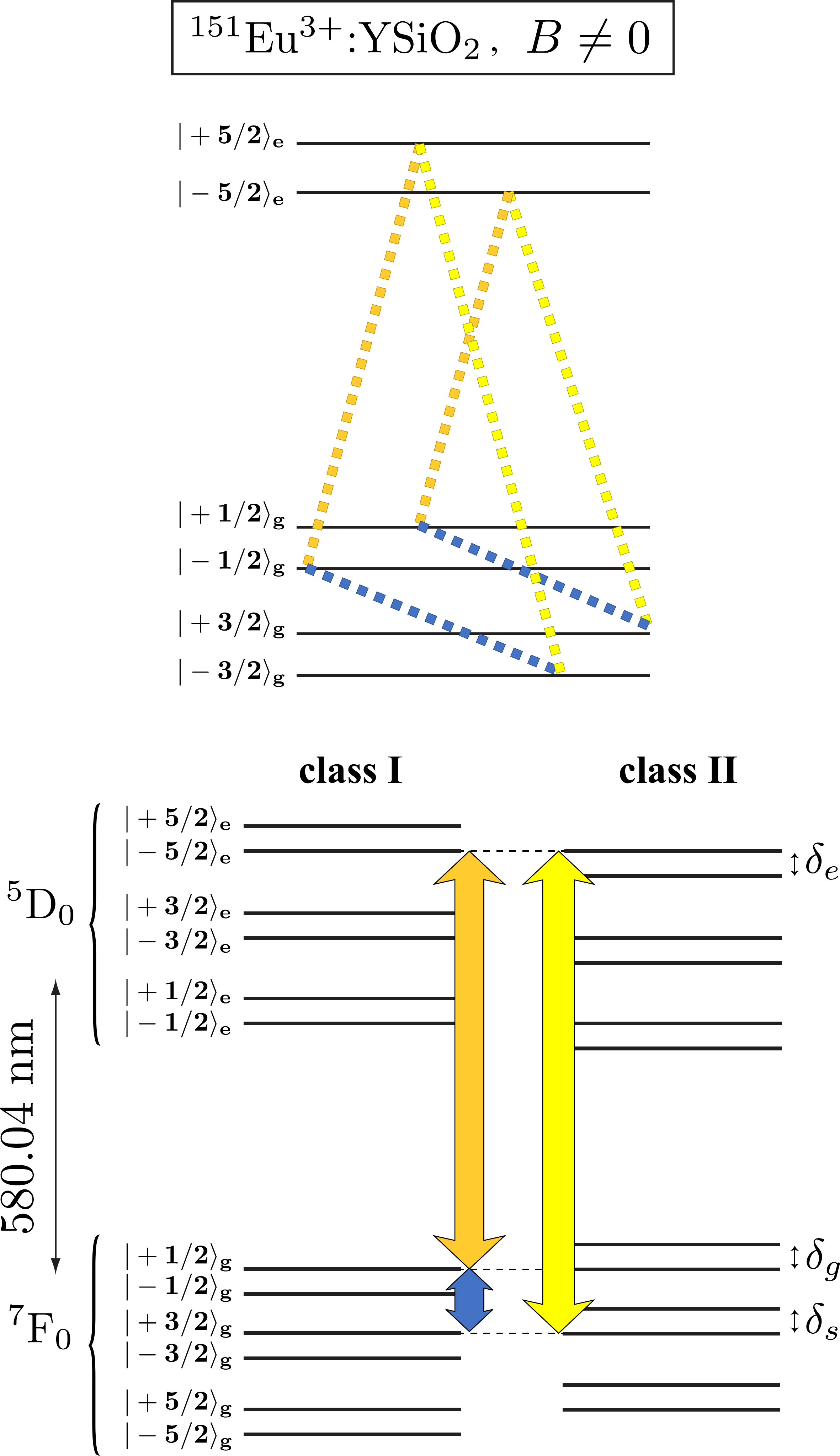}
		\caption{\label{FIG:twoclasses}In the \textbf{upper} part we indicate all the transitions of an ion, that we may drive strongly within the experiment. They form two independent, closed Lambda systems. In the optical domain these transition are selected through the optical branching ratios - the spin transitions we select by frequency. De facto only one transition of each color will be driven for each given ion, such that every ion will belong to one of two possible classes that are sketched in the \textbf{lower} part.
		}
		
	\end{SCfigure}

Within this plane we choose an angle of $\phi$ = $\phi^{\rm mag}$ = 65$^{\circ}$ relative to the D$_1$-axis of the crystal. For this configuration two particularly convenient properties emerge. \newline
The first property is linked with the branching ratios. Applying a magnetic field will split each of the optical transitions into four, but for a certain range of angles (including $\phi^{\rm mag}$), two of the transitions are strongly preferred regarding their respective branching ratio (see figure \ref{FIG:branchingratios}(a), (b)).
In consequence, the strong transitions form two independent Lambda systems. Each Zeeman state in the $\ket{1/2}_g$ manifold is connected to exactly one Zeeman state in the $\ket{5/2}_e$ manifold which in turn is connected to exactly one Zeeman state in the $\ket{3/2}_g$ manifold, as illustrated in figure \ref{FIG:twoclasses} (upper part).
	
	The other particularity at this angle is that the ground state splitting $\delta_g$ is equal to the storage state splitting $\delta_s$ (see figure \ref{FIG:exp_scheme} and figure \ref{FIG:twoclasses}(a)). This means that two of the four transitions  $\ket{1/2}_g\leftrightarrow\ket{3/2}_g$ are degenerate in frequency and can be driven simultaneously in an efficient manner with the RF field, while the other two are significantly detuned and weaker (figure \ref{FIG:branchingratios}(c)). Incidentally, the former two transitions close the two independent optical Lambda systems we address in our experiment (see figure \ref{FIG:twoclasses}).
	
The optical fields in the experiment have a bandwidth that is smaller than all of the relevant Zeeman splittings. As illustrated in figure \ref{FIG:twoclasses}, the light addresses only one single Lambda system at a time for any given ion. The inhomogeneous broadening of the respective transition, on the other hand, is much larger than the Zeeman splittings. Consequently, for some ions we are resonant with one of the strong lambda systems and for other ions the other one - two different \textit{classes} of ions are being addressed, as shown in the lower part of figure \ref{FIG:twoclasses}. Making use of these two classes simultaneously ensures that we do not lose any optical depth compared to the zero bias field scenario. Further, the radio frequency of the respective relevant spin transition is the same for both classes. Working with a bias field at angle $\phi^{\rm mag}$ allows for profiting from a significantly increased spin storage time without sacrificing optical depth or increasing the bandwidth of spins that we need to manipulate.

	
	\begin{figure*}[ht!]
\includegraphics[width=0.9\textwidth]{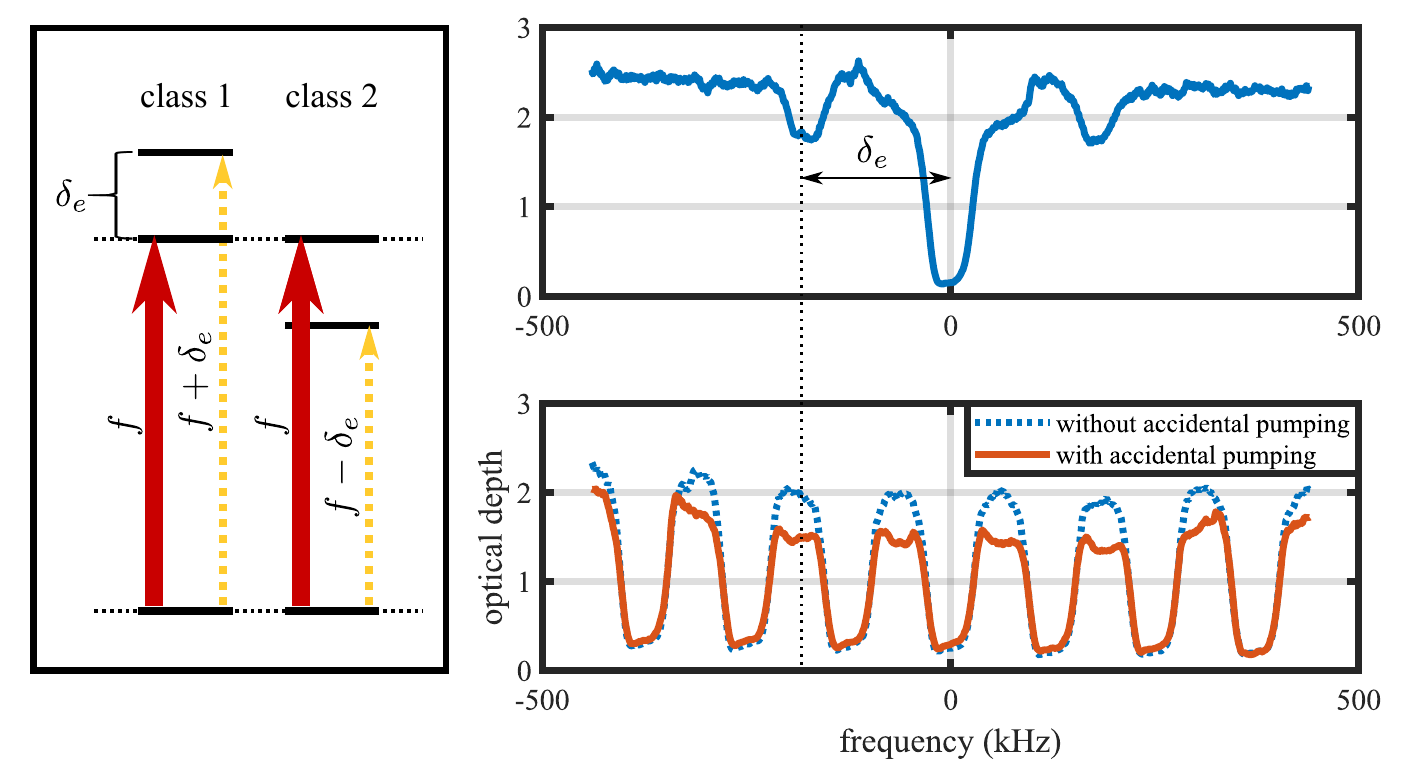}
\caption{\label{FIG:degradedcomb} 
\textbf{left:} In an inhomogeneously broadened ensemble with a split excited state optical pumping at a frequency $f$ not only produces increased transparency at $f$, but also at the frequencies $f+\delta_e$ and $f-\delta_e$ - the so called side-holes. \textbf{upper right:} Hole-burning spectrum in our system at a bias field of 10~mT (field orientation as in main text) with visible side-holes at $\pm$200~kHz. \textbf{lower right:} Comb that is deteriorated by the side-holes (red solid line) vs a comb that is not affected by the side-holes (blue dotted line}.
\end{figure*}	

	\subsection{Additional constraints for working with the magnetic bias field}	
	 While operating the experiment in this particular configuration is favorable in several regards, it does require the fulfillment of three additional constraints compared to operation at zero bias field.
	\\
The first two are connected to the fact that we want to address the two central nuclear transitions at 65$^\circ$  in figure \ref{FIG:branchingratios}(c), but not the two crossed ones. Any ion that is transferred to the wrong Zeeman state will no longer contribute to the collective re-emission of the ensemble - either because it is no longer in resonance with the control field, or because it acquires an additional phase relative to the rest of the ensemble.\\
	Selective transfer requires that the transition are well resolved - in other words the bias field must be sufficiently strong such that the undesired crossed transitions are much more separated from the central ones than the inhomogeneous linewidth of the individual transitions. \\
	Even in the case that these transitions are well separated, we could still drive them off-resonantly. In order to exclude this possibility, the detuning of our RF-field with regard to the unwanted transitions must be much bigger than the Rabi-frequency of the driving field, which corresponds to constraint (\ref{condrabi}) that we mention in the body of the article.
	\\
	 The third constraint is related to the atomic frequency comb preparation - while the contribution of the two weak optical transitions can be ignored in the storage process itself, this is not true for the AFC preparation. The comb is prepared by frequency selectively removing ions from the transition by optically pumping them into an auxiliary state that is not resonant to any of the light of the experimental sequence. In order to maximize the efficiency we apply many cycles of optical excitation and relaxation, meaning that even ions that are only resonant with a weak optical transition may be removed from the absorption. By this process, ions that otherwise might be contributing to the AFC are removed and therefore the optical depth (and consequently the efficiency of the memory) is reduced. The loss is associated with spectral side-holes that occur for optical pumping in inhomogeneous ensembles with split excited states (see figure \ref{FIG:degradedcomb}). This means that in our experiment accidental removal takes place when the transitions to both Zeeman states in the $\ket{5/2}_e$ manifold are within the memory bandwidth. In other words, accidental removal of ions can be avoided if the memory bandwidth is smaller than the Zeeman splitting $\delta_e$ in $\ket{5/2}_e$, as shown with constraint (\ref{condBWAFC}). In the case of our system the excited state splitting at 15~mT is approximately 300~kHz while the memory bandwidth is 160~kHz - so for our experiment the constraint is well-fulfilled.\\
	Alternatively, this detrimental effect could be avoided if the comb periodicity is matched to the excited state splitting. If the excited state splitting is a multiple of the comb periodicity, the position of the side-holes coincide with the position of neighboring holes, meaning that the ions that are removed over a weak transition are the ones that are supposed to be removed anyway so that there is no accidental loss of optical depth through this process.
	
		\subsection{Discussion of the storage decay curves for the XX sequence}
	As discussed in Sec. \ref{subseq:tau_dep}, the pulse area error model predicts a Gaussian decay of the output signal, while the OU spectral diffusion model predicts an exponential decay. Therefore we expect that for the XX sequence the storage decay curves would go from being more Gaussian to exponential as the pulse separation $\tau$ is increased. In such a situation one can use a stretched exponential as a model function, see Eq. \ref{EQ:Mims}. In addition, from the numerical calculations of the effect of pulse area errors presented in \cite{ZambriniCruzeiro2016}, one can also expect that the XX sequence produces decay curves with oscillations around a stationary value at long delays. A possible model function for the memory efficiency decay function for the XX sequence is then a stretched exponential with an offset:

	\begin{equation}
	\eta(t)=\eta_0\exp\left[-2\left(\frac{t}{T_2}\right)^\alpha\right]+c,
	\label{EQ:Mims_offs}
	\end{equation}	 
	which can describe both exponential ($\alpha=1$) and Gaussian ($\alpha=2$) behavior, as well as a smooth transition in-between the two.
	
	\begin{figure}[ht!]
		\includegraphics[width=1\textwidth]{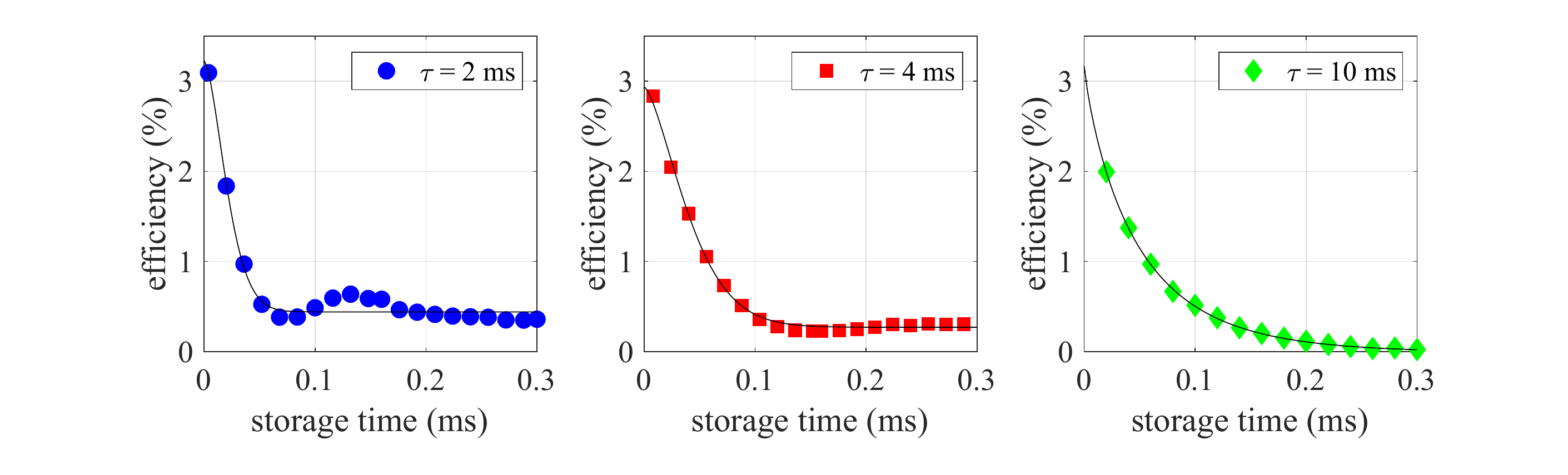}
		\caption{\label{FIG:Mims} Examples of three memory efficiency curves recorded using the XX sequence. All curves were fitted to the stretched exponential model with an offset, see Eq. \ref{EQ:Mims_offs}. The revivals for lower values of $\tau$ have been predicted in \cite{ZambriniCruzeiro2016} in the case of pulse area error, but are not considered in our very simple fitting model. The fitted $\alpha$ parameters of the stretched exponential can be found in Table \ref{TAB:Mims}.}
	\end{figure}
	
		If the efficiency decay curves of the XX sequence are fitted with the stretched exponential, we find that $\alpha$ tends towards larger values for shorter pulse separations $\tau$ and approaches one as we move to longer  $\tau$ (see Figure \ref{FIG:Mims} and Table \ref{TAB:Mims}). This is qualitatively in agreement with the theoretical modeling that we suggest for those two asymptotic regimes. However, as seen in Table \ref{TAB:Mims} the error bar of the fitted $\alpha$ parameter is particularly high where a more Gaussian decay is expected. In general the stretched exponential results in larger uncertainties in all estimated parameters, as an additional parameter is added to the fitting algorithm.
	
	\begin{table}[ht!]
			 
		\begin{center}
			{\small
				\begin{tabular}{ |c||c|c|c|c|c|c|c|c| } 
					\hline
					$\boldsymbol{\mathrm{\tau}}$ & 1.5 ms & 2 ms & 2.5 ms & 3 ms & 4 ms & 6 ms & 8 ms & 10 ms\\ 
					\hline 
					$\boldsymbol{\alpha}$ & 1.6$\pm$0.5 & 1.6$\pm$0.3 & 1.6$\pm$0.3 & 1.7$\pm$0.2 & 1.5$\pm$0.1 & 1.11$\pm$0.03 & 1.04$\pm$0.06 & 0.85$\pm$0.04 \\ 
					\hline 
				\end{tabular}
			}
		\end{center}
		\caption{\label{TAB:Mims} Dependence of the stretching coefficient $\alpha$ on the pulse separation $\tau$ in the XX dynamical decoupling scheme.  }
	\end{table}

	We also studied the robustness of the estimation of the decay constant using an exponential, stretched exponential, or a Gaussian decay. The fitted coherence time for the XX sequence was rather independent of the choice in model, the mean deviation between the models being at most 6\%. As the XY-4 and XY-8 sequences could be best fitted to an exponential function, which is expected due to the low impact of pulse errors for those sequences, we decided to fit all decay curves using an exponential function. We emphasize that the main conclusions presented in Sec. \ref{subseq:tau_dep} does not depend on the choice of the fit model.

\section*{References}
\bibliography{RefPD}
\bibliographystyle{iopart-num}

\end{document}